\documentclass[pdflatex,sn-mathphys-num]{sn-jnl}

\usepackage{graphicx}%
\usepackage{multirow}%
\usepackage{amsmath,amssymb,amsfonts}%
\usepackage{amsthm}%
\usepackage{mathrsfs}%
\usepackage[title]{appendix}%
\usepackage{xcolor}%
\usepackage{textcomp}%
\usepackage{manyfoot}%
\usepackage{booktabs}%
\usepackage{algorithm}%
\usepackage{algorithmicx}%
\usepackage{algpseudocode}%
\usepackage{listings}%
\usepackage{tikz,xcolor,hyperref,xurl}
\usepackage{comment}
\usepackage{subcaption}
\usepackage{tabularx}

\theoremstyle{thmstyleone}%
\theoremstyle{thmstyletwo}%

\theoremstyle{thmstylethree}%

\begin{document}

\title[Article Title]{The Dual Personas of Social Media Bots}


%
\author[1]{\fnm{Lynnette Hui Xian} \sur{Ng}}\email{huixiann@cs.cmu.edu}
\author[1]{\fnm{Kathleen M.} \sur{Carley}}\email{carley@cs.cmu.edu}

\affil[1]{\orgdiv{Software and Societal Systems Department}, \orgname{Carnegie Mellon University}, \orgaddress{\street{5000 Forbes Ave}, \city{Pittsburgh}, \postcode{15213}, \state{PA}, \country{USA}}}


\abstract{Social media bots are AI agents that participate in online conversations. Most studies focus on the general bot and the malicious nature of these agents. However, bots have many different personas, each specialized towards a specific behavioral or content trait. Neither are bots singularly bad, because they are used for both good and bad information dissemination. In this article, we introduce fifteen agent personas of social media bots. These personas have two main categories: Content-Based Bot Persona and Behavior-Based Bot Persona. We also form yardsticks of the good-bad duality of the bots, elaborating on metrics of good and bad bot agents. Our work puts forth a guideline to inform bot detection regulation, emphasizing that policies should focus on how these agents are employed, rather than collectively terming bot agents as bad.
}

\keywords{}




\maketitle

\section{Introduction}
\label{intro}
Artificial Intelligent (AI) agents on social media are called ``bots". A social media bot is an automated \textit{user} that creates and distribute \textit{content}, and form or dissolve \textit{interactions} on social media platforms \cite{ng2025social}. This extensive set of mechanics gives rise to the complex species of social media bot agents that utilize the range of actions available on a social media platform. 

Bots play an important role in our information ecosystem, disseminating important information \cite{chang2021social}, bridging people-based and interest-based communities \cite{ng2023deflating}, performing digital marketing for e-commerce store fronts \cite{zafar2025digital}, and posting scheduled tweets \cite{aldayel2022characterizing} and announcements for busy politicians and influencers \cite{ng2024cyborgs}. The automation that the bot affords provides opportunities in user engagement and publicity creation through the social media platform.

At the same time, bots do ill to our information ecosystem. They spread disinformation \cite{ng2024assembling,aimeur2023fake,freelon2022black}, polarize communities \cite{lu2024agents}, and proliferate propaganda \cite{marigliano2024analyzing,tardelli2024multifaceted}. Such online actions can result in harmful offline consequences such as vaccine hesitancy \cite{blane2023analyzing}, distributing of white supremacists flyers \cite{diab2023online}, or riots \cite{woolley2016social}. Bots are generally regarded as malicious agents because of the societal consequences of these offline disruptions. 

The bot agent is of importance to social cybersecurity research \cite{carley2020social}, a research area to characterize cyber-mediated changes and strengthen the social cyber-infrastructure. Good bots reinforce the social cyber infrastructure, particularly the social media platforms, by performing fact checking tasks \cite{nakov2021automated} or by amplifying government information \cite{he2024platform}. Bad bots undermine the social cyber infrastructure by creating chaos and sowing discord \cite{o2022automated}. 

The importance of bot agents in the health of social media platforms results in a long string of studies and analysis \cite{al2020bots,ng2021bot,tardelli2024multifaceted,beskow2018bot,shao2018spread}. Unfortunately, most studies focus on the malicious ones that cause harm to the online ecosystem through disinformation dissemination \cite{aimeur2023fake,shao2018spread} and information manipulation \cite{ng2022pro,lu2024agents,gambini2024anatomy}, and neglect to examine the good bots that quietly work behind the scenes to preserve social cyber health by performing activities such as correcting false information \cite{tsvetkova2017even}. Moreover, these studies typically group all detected bots into a single category to be treated similarly. Bot agents have different personas afforded to them based on their programming, and therefore should not be a singular collective when examined, because each persona can affect the information environment differently. The emphasis on negative impact of bot accounts has created a blind spot in our understanding of bot ecosystems. To that end, this article catalogues a set of bot personas and the goodness of their use.

This article unpacks the complexity of social media bots into fifteen unique personas. Each persona has dual use, a good side and a bad side. This categorization aids our understanding of the agents that operate in the social media space, and provides a frame to analyze how these agents can influence our online conversations. The scale and the speed at which these bots can operate magnifies their potential impact \cite{ferrara2016rise}, whether for good or for bad, thus there is a need for better awareness of the types and activities of the bot agents. Our framework can inform regulatory decisions surrounding bot agents and eventually create a better social cyberspace.

\section{The Personas of the Social Media Bot}
Bot agents have a diverse set of programming which to carry out their varied tasks on social media platforms. 
To date, we have identified fifteen different personas of social media bots, characterized by their behavior and content. These personas are created through inductive codes based on a literature survey, and were discussed extensively among the authors. 

In terms of analysis of the bot agent, past work have often relied on a binary classification to differentiate bots from humans. This typically involves machine learning classifiers that are trained on existing human annotations to provide the probability of whether a user is a bot or human. Common algorithms that are used to perform this social bot detection are: BotOrNot \cite{davis2016botornot}, DeBot \cite{chavoshi2016debot}, Botometer \cite{10.1145/3340531.3412698}, BotHunter \cite{beskow2018bot}, BotBuster \cite{ng2023botbuster} and TwiBot \cite{feng2022twibot}. These algorithms make use of hundreds to thousands of user-based and content-based features such as periodicity of postings, user metadata, content cues, to differentiate between bot and human.

Beyond a binary bot/human classification, more nuanced classifications of bot agents have been developed. Some typologies focus solely on one of the three pillars that describe a social media user -- the content or interaction mechanics. The Center of Naval Analysis binned bots into their content activity: distributing, amplifying, distorting or hijacking content, and flooding conversations or fracturing conversations \cite{mcbride2020social}. Another profiling method relies solely on the content, grouping bots into categories based on the amount of content (i.e., none, some, a lot) and content type (i.e., malicious or not) put out by the bot \cite{oentaryo2016profiling}. In this profiling, broadcast bots put out some information while spam bots post a voluminous amount of malicious information. \cite{abokhodair2015dissecting} differentiated bots by whether they put out original content or primarily retweeted content.

Another set of bot taxonomy rely on interaction mechanics used. Accounts that constantly retweet other accounts are known as retweeting bots \cite{elmas2022characterizing}, accounts that follow a lot of other accounts are known as fake-follower bots, and accounts that frequently spread malicious, phishing or scam content are are known as social spam bots, which are identified through large numbers of unrelated replies, user tags (@mentions), hashtags or duplicate content \cite{chu2012detecting}. 

Other bot taxonomies base their classifications only malicious bot agents. \cite{jamison2019malicious} defined bot personas by the impact of their presence in health discourse, classifying spam bots, content polluters and cyborgs. Similarly, \cite{lee2011seven} focused on malicious bots, identifying four types of bot personas based on the URLs shared and the follower-following network dynamics: duplicate spammers, duplicate @ spammers, malicious promoters and friend infiltrators. These typologies are constructed based on anecdotal evidence of bad bots, and need to be expanded towards the behaviors of good bots for a comprehensive characterization. 

While many past work have derived labels of types of bots, they do not have a cohesive definition. Much of these past work rely on the subjective process of human labeling to identify groups of bot types before developing a machine learning model to identify differentiating features of these sets of accounts. This process can result in a broad-based typology, such as that developed by \cite{mbona2023classifying}, which  states that all bots run automated tasks, but the difference between benign and malicious bots is that malicious bots perform malicious activities.
Further, the manual labeling process is time and resource intensive and is not scalable towards classifying tens of thousands of bots.
We must therefore classify bot personas with a consistent frame and develop an automated classification scheme. In this article, we adopt the user-content-interaction frame that describes social media users to classify different types of bots. Our bot persona classification breaks down the frame into measurable metrics, that can facilitate automated classification.  

We surveyed the literature for the analysis of bot activities to build our taxonomy. We annotated the bot activities in the user-content-interaction frame. Then we derived bot personas that embodied the mechanics employed. These bot personas fall into two big categories: bot personas by behavior and bot personas by content. We characterize each of these personas in terms of the first-principles definitions of a bot, laying out the user-content-interaction characteristics of each bot persona. The fifteen personas and their definitions are reflected in \autoref{tab:personas_1} and \autoref{tab:personas_2}. The breakdown of properties of personas served as heuristics for identification of each persona. Such heuristics can also be used for generating the personas, as observed in the AuraSight scenario \cite{ng2025aurasight}, where simulated social media data were generated.

\begin{table}[h!]
    \centering
    \begin{tabular}{p{2.5cm}p{3cm}p{3cm}p{3cm}p{3cm}}
    \hline
         \textbf{Persona} & \textbf{Definitions} & \multicolumn{3}{c}{\textbf{Properties}} \\ \hline 
         ~ & ~ & \textbf{User} & \textbf{Content} & \textbf{Interaction} \\ \hline 
         \multicolumn{4}{l}{\textbf{Behavior-Based Bot Persona}} \\ \hline
         \textbf{Social Influence Bot} & Attempt to influence other users' perceptions & & High usage of emotional and BEND cues\textsuperscript{1} & Excessive replies \\ \hline
         \textbf{Amplifier Bot} & Boost narrative themes and manufacture support & ~ & ~ & Excessive sharing (i.e., retweets, shares) pattern \\ \hline
         \textbf{Cyborgs} & Exhibit both human- and bot-like activity & Frequent changes in bot classification with high change in bot probability score & & \\ \hline
         \textbf{Bridging Bot} & Connects groups of users & High Bridge score from the BEND metric & & Frequently tag multiple people from different social identity groups, or straddle between multiple network clusters \\ \hline  
         \textbf{Repeater Bot} & Repeats posts, sometimes keeping most of the content the same & ~ & Frequent posts with similar texts & \\ \hline
         \textbf{Self-Declared Bot} & Outwardly declare themselves as a bot & Presence of word ``bot" in user information (i.e., username, screenname, description) & & \\ \hline
         \textbf{Synchronized Bot} & Coordinate with other bot accounts & High coordination index & & High number of bots that account is coordinating with \\ \hline
         \multicolumn{4}{l}{\textbf{Content-Based Bot Persona}} \\ \hline
         \textbf{Chaos Bot} & Sow chaos and discord & Higher than average BEND values$^{1}$ & & \\ \hline 
         \textbf{Announcer Bot} & Announce information & & Periodic posting patterns, or posting with templates & \\ \hline 
         \textbf{Content Generation Bot}  & Generate content for the online ecosystem & & Majority of posts are original content rather than shared content & \\ \hline 
         \textbf{Information Correction Bot} & Correct information & & Presence of negation & References to fact checking website\\ \hline 
         \textbf{Genre Specific Bot}  & Posts mostly on a singular topic & & Majority of the posts on a single topic frame & \\ \hline 
         \textbf{Conversational Bot} & Carry out a conversation with other users & & Content has higher than average reading difficulty and varied use of topic frames & \\ \hline 
         \textbf{Engagement Generation Bot} & Incite engagement from other users & & Use of emotional cues and high emotional valence & \\ \hline 
         \textbf{News Bot} & Post news updates & Presence of the word ``news" in user metadata & Majority of posts are news headlines & \\ \hline 
    \end{tabular}
    \caption{Personas of social media bots (By Behavior). \textsuperscript{1}BEND cues are signals of information maneuver tactics, derived from \cite{ng2025social}.}
    \label{tab:personas_1}
\end{table}

Finally, we ran these heuristics on a subset of data taken from the 2020-2021 coronavirus vaccine discussions on X. These discussions were based on the conversations that took place the week before the coronavirus vaccine rollout (1 December - 7 Decemeber, 2020). The data was collected through the Twitter V1 Streaming API and filtered for vaccination terms, and consisted of 1,076,734 tweets from 580,135 unique users. We urge the reader to refer to the original article for the data collection details \cite{blane2023analyzing}. We applied the heuristics from \autoref{tab:personas_1} and \autoref{tab:personas_2} on the data, and summarize examples of tweets from each type of bot in \autoref{tab:examples_1} and \autoref{tab:examples_2}.

\begin{table}[h!]
    \centering
    \begin{tabular}{p{3cm}p{10cm}}
    \hline
         \textbf{Persona} & \textbf{Illustrated examples}  \\ \hline 
         \multicolumn{2}{l}{\textbf{Content-Based Bot Persona}} \\ \hline
         \textbf{Chaos Bot} & From the same user: ``DON'T TAKE IT PEOPLE! I spoke to old lady today and her daughter has told her she cant see grand daughter  without having vaccine WHAT A XXXXXXX BITCH", ``Former Pfizer Science Officer Reveals Great COVID-19 Scam" \\ \hline 
         \textbf{Announcer Bot} & @GlobalPandemic begins each tweet with ``ALERT:". For example, ``ALERT: Novavax, Maryland biotech firm, completes enrollment for COVID-19 vaccine trials [...]", ``ALERT: CDC-convened committee votes to add nursing home residents to first phase of COVID-19 vaccine access [...]", ``ALERT: U.K. becomes first country to approve Pfizer/BioNTech COVID-19 vaccine [...]" \\ \hline 
         \textbf{Content Generation Bot}  & @quant\_data tweets original tweets about the financial markets ``Moderna shares are trading higher on optimism surrounding the company's COVID-19 vaccine candidate.  Related Tickers: \$MRNA [URL]", ``Carnival shares are trading higher as positive COVID-19 vaccine news in recent weeks has boosted investor sentiment and lifted cruise sector recovery hopes.  Related Tickers: \$CCL [URL]" \\ \hline 
         \textbf{Information Correction Bot} & @FactCheck tweets: ``Terrific questions - thank you. I have put question 1 to several experts and they have told me that they generally expect side-effects from a vaccine injection to show up quickly. Delayed reactions are unusual. (1)" \\ \hline 
         \textbf{Genre Specific Bot} & @IndiaTrendDaily mostly talks about latest gear for sales. Example tweets: ``DELL Inspiron 16inch HD Laptop [URL]", ``HP 1.6-inch Laptop [URL]" \\ \hline 
         \textbf{Conversational Bot} & @CuddleTalk: ``@greta Greta, anyone taking an mRNA COVID-19 vaccine that modifies one's mRNA to produce the spike proteins on a COVID-19 [...]", ``@GiannoCaldwell Why would anyone take a COVID-19 vaccine that modifies their mRNA to produce the spike proteins on the outside of the COVID-19 cell to produce an immune response?"  \\ \hline 
         \textbf{Engagement Generation Bot}  & @DonnaYouDC tries to get readers to think about the anti-vaccination arguments: ``And if the \#pharma \#biotech, pharmacy \& distributor execs are "irritated at pressure from the White House to attend an event they perceive to be largely political," they certainly don't have to attend. So, will they anyway?  \$PFE \$MRNA \$FDX \$UPS \$MCK \$WBA \$CVS \#COVID19 \#vaccines", ``Why would \#pharma \#biotech, etc. want Trump to take credit for their work?  \$PFE \$MRNA \$FDX \$UPS \$MCK \$WBA \$CVS \#COVID19 \#vaccines" \\ \hline 
         \textbf{News Bot} & Example users: @dwatchnews\_asia, @novasocialnews, @HealthNewsFL \newline Example tweets: ``Pfizer seeks emergency approval for COVID-19 vaccine in India, media say [URL]"; ``China prepares large-scale rollout of coronavirus vaccines (from @AP) [URL]" \\ \hline 
    \end{tabular}
    \caption{Examples of tweets by personas of social media bots (By Content).}
    \label{tab:examples_1}
\end{table}

\begin{table}[h!]
    \centering
    \begin{tabular}{p{3cm}p{10cm}}
    \hline
         \textbf{Persona} & \textbf{Illustrated examples}  \\ \hline 
         \multicolumn{2}{l}{\textbf{Behavior-Based Bot Persona}} \\ \hline
         \textbf{Social Influence Bot} & ``STOP THE BLOODTHIRSTY LUCIFARIAN GLOBALISTS AND YOUR PLANDEMIC WHICH IS REALLY A...\#LucifarianNewWorldOrderTakeover \#TheGreatEvilReset \#Agenda2030 \#Agenda21 \#Depopulation \#WeDoNotConsentCVVaccine2020 \#JesusIsLord \#NotBillGates \#GodWins" \\ \hline
         \textbf{Amplifier Bot} & Tweets from the same user @CEPIvaccines: ``RT @UN\_PGA: We need to join forces to ensure \#Vaccines4All. Fair and equitable access to \#COVID19 vaccines is both the right thing to do \& ...", ``RT @gavi: The Gavi \#COVAX Advance Market Commitment (AMC) is an innovative financial mechanism designed to ensure that the world’s poorest ...", ``RT @gavi: \#COVAX is founded on the principle of equitable access because in the \#COVID19 vaccine race, nobody wins unless everyone wins. ..."  \\ \hline
         \textbf{Cyborgs} & @IAM\_\_Network uses several platforms to tweet (Blog2Social, IAMBLOG2TWITTER, Microsoft Power Platform) \\ \hline
         \textbf{Bridging Bot} & Example tweets: ``@LongJonson2 @JimmyJGreen @wbz If interested I wrote up exactly how these vaccines work. This doesn’t change your DNA at all. It utilizes the ribosomes in your cells to read an mRNA strand to create the protein. This is something our cells do naturally every day https://t.co/fdRTIK7Xie" ; ``@RenTradewind @entobird @JUNIUS\_64 @ROCKETDRAG @MarziManed @CatsCavern Molecular Biologist here! I study Viral Diseases, Viral Immunity and Pathogenesis for a living at the National Institute of Allergy and Infectious Diseases. Current research efforts have been dedicated to the Coronavirus vaccine. \#FurriesinSTEM [URL]"  \\ \hline  
         \textbf{Repeater Bot} & @ElsaKekita repeats the same tweet more than 3 times ``La vacuna de Moderna otorga al menos 3 meses de inmunidad frente el coronavirus - Infobae [URL]"  \\ \hline
         \textbf{Self-Declared Bot} & Example Users: @CoronaUpdateBot, @Bot\_Corona\_V, @GothBotAlice\\ \hline
         \textbf{Synchronized Bot} & @TheNewsGob, @OnlyFilmMedia, @VintageVideoPod, @newsbreakApp frequently tag each other  \\ \hline
    \end{tabular}
    \caption{Examples of tweets by personas of social media bots (By Behavior).}
    \label{tab:examples_2}
\end{table}

\section{The Duality of Social Media Bot}
A bot can exhibit either good or bad qualities depending on its environmental conditions, which includes its narrative and its social network. Whether the bot is good or bad depends on its environment, such as its usage, the content it spreads, or the social network interactions it forms. In other words, it is not the bot technology that is good or bad, it is the way the technology is employed by the human controllers that makes it good or bad.

Bot detection algorithms \cite{davis2016botornot,chavoshi2016debot,beskow2018bot,10.1145/3340531.3412698,ng2023botbuster,feng2022twibot} are able to reliably differentiate whether a user is an automated agent or not, but these algorithms fail to differentiate between beneficial and harmful bots. Should these algorithms be plainly implemented on social media platforms, it can result in the removal of legitimate bot accounts while malicious networks remain operational in the social cyberspace \cite{ferrara2020characterizing}. To this problem, we develop a metric to differentiate between good and bad bot behavior. Our metrics of good and bad bots are presented in \autoref{tab:duality_behavior_1} and \autoref{tab:duality_behavior_2}. These metrics differentiate the goodness of the bot by two axes: the content the bot puts out and the interaction the bot has with other users.

Good bots spread good content and seek to promote positive emotions (e.g., happy), while bad bots spread bad content, sometimes with links that point to disinformation sites, and promote negative emotions (e.g., anger, fear). Good bots also tend to use more positive information maneuver techniques, which can be measured with the BEND framework \cite{carley2020social}. The BEND framework is a set of 16 information maneuvers that manipulate the social network interaction structure. Good bots use constructive maneuvers, such as back and build to build positive communities and back affirmative narratives, creating an inclusive social cyberspace. Bad bots use destructive maneuvers, such as dismiss and distort to disperse groups and misrepresent facts, creating an exclusive social cyberspace.

Bots used in crisis communication exemplify the positive impact that these AI agents can have. For example, announcer bots are used to alert on social media on impending earthquakes \cite{avvenuti2014ears}. In the same scenario, chaos bots use social media as a distributed sensor system to aggregate and make sense of the ground-zero situation during an earthquake \cite{crooks2013earthquake}. Announcer bots are also used for weather alerts, to provide timely forecasts and warnings, which gives people time to prepare for adverse conditions \cite{hofeditz2019meaningful}.

Genre-specific bots in the political realm demonstrate the negative impact of bot agents. Synchronized bots have been deployed to artificially inflate support for polit,ical candidates \cite{ng2024tiny,woolley2016automating}, and push specific tweets from  narratives in a coordinated fashion \cite{jacobs2023tracking,ng2023combined}. Overall, bots manipulate the public discourse \cite{ng2022pro}, and are regarded to have affected election outcomes \cite{badawy2018analyzing}.

The goodness of a bot is not static throughout its lifespan, and can change over time. This happens when bots begin out as good bots that post benign content to gain traction, before posting bad content. For example, \cite{ng2021bot} found bots that were originally food accounts before evolving into bots that call for violent action during the Indian-Pakistan-Kashmir conflict in 2021. Such bots that turn from good to bad are likely to be good agents in between their targeted events in order to maintain their account, or are hijacked accounts \cite{elmas2022role}. In the former case, the bots may cycle through good and bad behavior, and will require constant monitoring.

\begin{table}[h!]
    \centering
    \begin{tabular}{p{2cm}p{2.5cm}p{2.5cm}p{2.5cm}p{2.5cm}}
    \hline
        ~ & \multicolumn{2}{c}{\textbf{Good Agent}} & \multicolumn{2}{c}{\textbf{Bad Agent}} \\ \hline
        ~  &  \textbf{Content} & \textbf{Interaction} & \textbf{Content} & \textbf{Interaction} \\ \hline
        \textbf{Social Influence Bot} & good content, positive emotion cues & bad content, negative emotion cues & cites disinformation sources \\ \hline 
        \textbf{Amplifier Bot} & good content & high number of retweets & bad content & cites disinformation source, high number of retweets \\ \hline
        \textbf{Cyborgs} & good content & negative change in message tone & bad content & cites disinformation source \\ \hline 
        \textbf{Bridging Bot} & good content & & bad content & cites disinformation source \\ \hline
        \textbf{Repeater Bot} & good content, posts same message & ~ & bad content, posts same messages & cites disinformation source \\ \hline 
        \textbf{Self-Declared Bot} & good content & bad content & cites disinformation source \\ \hline 
        \textbf{Synchronized Bot} & good content & & bad content & cite disinformation source \\ \hline 
    \end{tabular}
    \caption{Duality of social media bots (By Behavior)}
    \label{tab:duality_behavior_1}
\end{table}

\begin{table}[h!]
    \centering
    \begin{tabular}{p{2cm}p{2.5cm}p{2.5cm}p{2.5cm}p{2.5cm}}
    \hline
        ~ & \multicolumn{2}{c}{\textbf{Good Agent}} & \multicolumn{2}{c}{\textbf{Bad Agent}} \\ \hline
        ~  & \textbf{Content} & \textbf{Interaction} & \textbf{Content} & \textbf{Interaction} \\ \hline
        \textbf{Chaos Bot} & good content & constructive BEND maneuvers & bad content & destructive BEND maneuvers, cites disinformation source \\ \hline 
        \textbf{Announcer Bot, Content Generation Bot, Information Correction Bot, Genre Specific Bot, Conversational Bot, Engagement Generation Bot, News Bot } & good content & & bad content & cites disinformation source \\ \hline 
    \end{tabular}
    \caption{Duality of social media bots (By Content)}
    \label{tab:duality_behavior_2}
\end{table}

\section{Dual Personas of Social Media Bots}
Here we discuss the dual personas of the social media bot as observed in literature. These studies are presented in Table~\ref{tab:bot_usage_behav} and ~\ref{tab:bot_usage_content}. Much of the bot analysis literature focuses on the content of bot agents, observing many content-based bot personas, such as the content generation bot and the genre specific bot. The bot personas found in literature are mostly situated with bad content, such as opinion manipulation \cite{chen2022social}, inciting online polarization \cite{danaditya2022curious} and disinformation spread \cite{aimeur2023fake,shao2018spread}.
In terms of behavior-based bot personas, the amplifier bot \cite{mckelvey2017computational,jacobs2023tracking,duan2022algorithmic} and the synchronized bot are most studied \cite{magelinski2022synchronized,tardelli2024multifaceted}, particularly in the context of political manipulation and polarization \cite{mckelvey2017computational,uyheng2020bot}.

A single bot can have multiple personas at once. The most common persona combination is a content-based persona and a behavior-based persona; such as an news amplifier bot. Genre-specific bots typically take on many different behaviors: a political genre-specific bot can take on a bridging bot persona that connects multiple political entities \cite{ng2023deflating}, a news bot persona that spreads political news \cite{ng2023deflating}, or an amplifier bot persona that disseminates political propaganda \cite{tardelli2024multifaceted}.  Many news bots are also synchronized bots, especially when they originate from the same parent company but are region-specific agents \cite{magelinski2022synchronized}. The synchronized 7News bots are news bots that disseminate news from the Australian network 7News. The parent account, @7News, typically originates tweets on global news, and the region-specific accounts like @7NewsBrisbane will retweet the original global news tweets in synchrony. These region-specific accounts are also responsible for tweeting region-related news, which are only broadcasted on the corresponding accounts.

A bot agent is not constrained to a single persona through its lifespan. The bot persona can change during its activity, sometimes flipping between personas. \cite{jacobs2023tracking} characterized bots that alternate between amplifier and repeater bots while communicating about election discourse. Cyborgs are used by famous and influential people, and are observed to take on announcer bot personas when broadcasting events like local community fundraising to a social influence bot when encouraging an action like taking up vaccination \cite{ng2024cyborgs}.

Similarly, the same type of bot persona can be both good and bad during their lifetime, embodying the duality of a bot. Announcer bots have been discovered to originally promote restaurants and later began  announcing calls for violent action \cite{ng2021bot}. Other bots simultaneously promote the government that they are affiliated with and discredit opposing ideas with false information \cite{marigliano2024analyzing}.

\begin{table}[h!]
    \centering
    \begin{tabular}{p{3cm}p{4cm}p{5cm}}
        \textbf{Persona} & \textbf{Usage for Good} & \textbf{Usage for Bad} \\ \hline
        \multicolumn{3}{l}{\textbf{Behavior-Based Bot Persona}} \\ \hline
        \textbf{Social Influence Bot} & change people's views towards vaccination (anti to pro) \cite{ng2022pro}, influence purchase intentions \cite{sindhu2024influence}& change people's views towards vaccination (pro to anti) \cite{ng2022pro}, opinion manipulation \cite{chen2022social}\\ \hline
        \textbf{Amplifier Bot} & amplifying government or regional news during crisis situations \cite{chang2022comparative} & amplify politically divisive narratives and narratives that target other countries\cite{jacobs2023tracking,duan2022algorithmic}, disseminate computational propaganda through ``flooding the zone" technique\cite{o2022automated}, amplify hate speech, amplify the spread of misinformation, inflate popularity of candidates during election season \cite{mckelvey2017computational}  \\ \hline
        \textbf{Cyborgs} & alleviate workload of social media managers, politicians and influencers \cite{ng2024cyborgs,gorwa2020unpacking} & trigger and initiate activism \cite{lotan2011arab,magelinski2022synchronized} \\ \hline
        \textbf{Bridging Bot} & political commentators that aggregate information across multiple parties \cite{ng2024tiny} & cross-cultural social marketing \cite{badawy2018analyzing}, information dissemination across groups for political manipulation \cite{ng2023deflating} \\ \hline
        \textbf{Repeater Bot} & constant product promotion and advertisement \cite{liu2017investigation} & constant sharing politically divisive narratives \cite{jacobs2023tracking}, spam content \cite{aldayel2022characterizing} \\ \hline 
        \textbf{Self-Declared Bot} & useful and creative bot accounts that label themselves as bots \cite{aldayel2022characterizing} & \\ \hline
        \textbf{Synchronized Bot} & synchronization in broadcasting news to region specific accounts \cite{magelinski2022synchronized}, raise citizen concerns on government actions \cite{ng2023you} & social botnets in political conflicts \cite{abokhodair2015dissecting}, participation in online influence campaigns \cite{khaund2021social} \\ \hline
    \end{tabular}
    \caption{Observations of the Duality of the Social Media Bot (By Behavior) }
    \label{tab:bot_usage_behav}
\end{table}

\begin{table}[h!]
    \centering
    \begin{tabular}{p{4cm}p{5cm}p{5cm}}
        \textbf{Persona} & \textbf{Usage for Good} & \textbf{Usage for Bad} \\ \hline
        \multicolumn{3}{l}{\textbf{Content-Based Bot Persona}} \\ \hline
        \textbf{Chaos Bot} & bring attention to citizen concerns like disrespect of muslim ideology \cite{ng2023you}, climate change \cite{lu2024agents}, organize resources during chaos in crisis \cite{himelein2021bots}, organize earthquake reporting from twitter \cite{avvenuti2014ears,crooks2013earthquake} & infiltration of an organization \cite{elyashar2014guided}, sow discord to polarize opinions \cite{lu2024agents} \\ \hline
        
        \textbf{Announcer Bot} & promotion of product launch \cite{liu2017investigation}, announcing locations to get help, e.g. vaccine, crisis \cite{chang2021social}, earthquakes \cite{avvenuti2014ears} & broadcasting propaganda messages \cite{arnaudo2017computational}, call for action against political incumbents \cite{ng2024assembling}, publicize protests locations and information \cite{ferrara2020covid} \\ \hline
        
        \textbf{Content Generation Bot} & warning for natural disasters \cite{haustein2016tweets}, crisis communication and emergency resource deployment \cite{hofeditz2019meaningful}, editing Wikipedia entries \cite{tsvetkova2017even} & generation of harmful racial and ideological content \cite{freelon2022black}, hijack conversations to create ideological polarization \cite{danaditya2022curious}, digital marketing for ecommerce \cite{zafar2025digital} \\ \hline
        
        \textbf{Information Correction Bot} & bots on Reddit moderate conversations \cite{he2024platform,jhaver2019human}, bots on Wikipedia prevent vandalism \cite{tsvetkova2017even} & correct true information into false ones \cite{ferrara2020characterizing} \\ \hline
        
        \textbf{Genre Specific Bot} & brand establishment and amplification \cite{ng2024cyborgs,uyheng2020bot}, digital campaigning \cite{ng2024tiny} through candidate promotion \cite{mckelvey2017computational}, religious prayer and worship generation \cite{ohman2019prayer} & political manipulation \cite{badawy2018analyzing}, use religion to influence audience \cite{ohman2019prayer}, radicalization and recruitment for extremist groups \cite{mantello2023automating,benigni2017online}, political accounts that systematically delete content \cite{aldayel2022characterizing} \\ \hline
        
        \textbf{Conversational Bot} & provide emotional support during stress \cite{piccolo2021chatbots} and grieve \cite{krueger2022communing}, provide help in frequently asked questions for e-commerce sites\cite{moriuchi2021engagement} & converses with racist and derogatory terms \cite{neff2016talking}, exacerbates stereotypes, and gender/ race divides \cite{wolf2017we} \\ \hline
        
        \textbf{Engagement Generation Bot} & provide humor \cite{veale2015twitter}, increase perceived product value and user engagement in e-commerce \cite{elsholz2019exploring} & artificially inflate engagement or popularity of malicious users, e.g. disinformation spreaders \cite{glenski2020user}, creation of viral engagement of bad content using botnet \cite{yeo2022models} \\ \hline
        
        \textbf{News Bot} & news production, dissemination and interaction with audiences in current media environment \cite{hong2020utilizing,lokot2016news}, active news promoters \cite{al2020bots}, curate target content from multiple information streams \cite{lokot2016news} & spread disinformation news and fake news \cite{aimeur2023fake,shao2018spread}, spread news through click baits to earn money \cite{al2020bots} \\ \hline
    \end{tabular}
    \caption{Observations of the Duality of the Social Media Bot (By Content) }
    \label{tab:bot_usage_content}
\end{table}

\section{Conclusion}
Social media bots are agents with a variety of personalities. Bots should not be treated collectively as an agent with a single personality type, because each personality impacts the online conversation differently. Each persona type has a dual use: for good or for bad. The goodness of a social media bot should be evaluated according to their content and interactions, rather than jointly flagging all bots as bad.

Our Dual Personality of Bots framework points to directions of future lines of research involving the distribution of information by different bot personas. Each persona distributes content with different interaction mechanics, and further studies should look at how the scope of content distribution and the engagement content differs with different mechanics. This leads to further investigation surrounding the distribution of persona types in different events and platforms, and catalog the bot personas that are typically active in different event types, e.g., politics, natural disasters. Such profiling will inform studies on the agents which are typically employed for the distribution and manipulation of information, and formulate communication strategies to deploy agents that leverage specific persona properties for the social good. For example, a government agency that seeks to publicize public health initiatives should use an announcer bot and schedule periodic, templated posts. 

Another key direction is operationalizing this framework by developing robust heuristics to detect these bot personas and classify their duality, such as was begun in \citep{ng2023deflating,jacobs2023tracking}. Such a development involves a multidisciplinary methodology that involves computational methods, network science and linguistic methods, to characterize the bot agent ecosystem. Algorithm development will allow efficient identification of bot personas within a large scale of data, aiding to downstream analysis. 

Further, understanding the dual personality of bots can inform content moderation and social media policy. In formulating policies and regulations, it is important that the complexity of the bot ecosystem is considered. Policies should not be formulated around what is used (i.e., whether a bot agent is deployed), but should examine how the agents are used. For example, a society that does not want disinformation to spread might ban an amplifier bot that spreads inaccurate information \cite{jacobs2023tracking}, but that same society might want to allow an amplifier bot that spreads public information such as access to clean water and food during a natural fire or hurricane disaster.

In conclusion, social media bot agents have varied personalities afforded to them. Each personality can be developed towards a good side and a bad side. Our Dual Persona framework refines the interpretation of the AI bot agent. By recognizing the nuanced roles and behaviors of different bot personas, we can develop more sophisticated detection methods, and improve policies that address specific challenges and leverage the capabilities presented by each persona. Our framework opens opportunities for academics, companies and policymakers to design interactions and regulations to create a positive online ecosystem where both humans and bots can exist symbiotically.

\section{Acknowledgments}
This material is based upon work supported by the Scalable Technologies for Social Cybersecurity, U.S. Army (W911NF20D0002), the Minerva-Multi-Level Models of Covert Online Information Campaigns, Office of Naval Research (N000142112765), the Threat Assessment Techniques for Digital Data, Office of Naval Research (N000142412414), and the MURI: Persuasion, Identity \& Morality in Social-Cyber Environments (N000142112749), Office of Naval Research. The views and conclusions contained in this document are those of the authors and should not be interpreted as representing official policies, either expressed or implied by the Office of Naval Research, U.S. Army or the U.S. government.

\newpage
\bibliography{biblio}


\begin{thebibliography}{84}
\ifx \bisbn   \undefined \def \bisbn  #1{ISBN #1}\fi
\ifx \binits  \undefined \def \binits#1{#1}\fi
\ifx \bauthor  \undefined \def \bauthor#1{#1}\fi
\ifx \batitle  \undefined \def \batitle#1{#1}\fi
\ifx \bjtitle  \undefined \def \bjtitle#1{#1}\fi
\ifx \bvolume  \undefined \def \bvolume#1{\textbf{#1}}\fi
\ifx \byear  \undefined \def \byear#1{#1}\fi
\ifx \bissue  \undefined \def \bissue#1{#1}\fi
\ifx \bfpage  \undefined \def \bfpage#1{#1}\fi
\ifx \blpage  \undefined \def \blpage #1{#1}\fi
\ifx \burl  \undefined \def \burl#1{\textsf{#1}}\fi
\ifx \doiurl  \undefined \def \doiurl#1{\url{https://doi.org/#1}}\fi
\ifx \betal  \undefined \def \betal{\textit{et al.}}\fi
\ifx \binstitute  \undefined \def \binstitute#1{#1}\fi
\ifx \binstitutionaled  \undefined \def \binstitutionaled#1{#1}\fi
\ifx \bctitle  \undefined \def \bctitle#1{#1}\fi
\ifx \beditor  \undefined \def \beditor#1{#1}\fi
\ifx \bpublisher  \undefined \def \bpublisher#1{#1}\fi
\ifx \bbtitle  \undefined \def \bbtitle#1{#1}\fi
\ifx \bedition  \undefined \def \bedition#1{#1}\fi
\ifx \bseriesno  \undefined \def \bseriesno#1{#1}\fi
\ifx \blocation  \undefined \def \blocation#1{#1}\fi
\ifx \bsertitle  \undefined \def \bsertitle#1{#1}\fi
\ifx \bsnm \undefined \def \bsnm#1{#1}\fi
\ifx \bsuffix \undefined \def \bsuffix#1{#1}\fi
\ifx \bparticle \undefined \def \bparticle#1{#1}\fi
\ifx \barticle \undefined \def \barticle#1{#1}\fi
\bibcommenthead
\ifx \bconfdate \undefined \def \bconfdate #1{#1}\fi
\ifx \botherref \undefined \def \botherref #1{#1}\fi
\ifx \url \undefined \def \url#1{\textsf{#1}}\fi
\ifx \bchapter \undefined \def \bchapter#1{#1}\fi
\ifx \bbook \undefined \def \bbook#1{#1}\fi
\ifx \bcomment \undefined \def \bcomment#1{#1}\fi
\ifx \oauthor \undefined \def \oauthor#1{#1}\fi
\ifx \citeauthoryear \undefined \def \citeauthoryear#1{#1}\fi
\ifx \endbibitem  \undefined \def \endbibitem {}\fi
\ifx \bconflocation  \undefined \def \bconflocation#1{#1}\fi
\ifx \arxivurl  \undefined \def \arxivurl#1{\textsf{#1}}\fi
\csname PreBibitemsHook\endcsname

\bibitem[\protect\citeauthoryear{Ng and Carley}{2025}]{ng2025social}
\begin{botherref}
\oauthor{\bsnm{Ng}, \binits{L.H.X.}},
\oauthor{\bsnm{Carley}, \binits{K.M.}}:
What is a social media bot? a global comparison of bot and human characteristics.
arXiv preprint arXiv:2501.00855
(2025)
\end{botherref}
\endbibitem

\bibitem[\protect\citeauthoryear{Chang et~al.}{2021}]{chang2021social}
\begin{bchapter}
\bauthor{\bsnm{Chang}, \binits{H.-C.H.}},
\bauthor{\bsnm{Chen}, \binits{E.}},
\bauthor{\bsnm{Zhang}, \binits{M.}},
\bauthor{\bsnm{Muric}, \binits{G.}},
\bauthor{\bsnm{Ferrara}, \binits{E.}}:
\bctitle{Social bots and social media manipulation in 2020: The year in review}.
In: \bbtitle{Handbook of Computational Social Science, Volume 1},
pp. \bfpage{304}--\blpage{323}.
\bpublisher{Routledge}, \blocation{???}
(\byear{2021})
\end{bchapter}
\endbibitem

\bibitem[\protect\citeauthoryear{Ng and Carley}{2023}]{ng2023deflating}
\begin{barticle}
\bauthor{\bsnm{Ng}, \binits{L.H.X.}},
\bauthor{\bsnm{Carley}, \binits{K.M.}}:
\batitle{Deflating the chinese balloon: types of twitter bots in us-china balloon incident}.
\bjtitle{EPJ Data Science}
\bvolume{12}(\bissue{1}),
\bfpage{63}
(\byear{2023})
\end{barticle}
\endbibitem

\bibitem[\protect\citeauthoryear{Zafar et~al.}{2025}]{zafar2025digital}
\begin{barticle}
\bauthor{\bsnm{Zafar}, \binits{H.}},
\bauthor{\bsnm{Siddiqui}, \binits{F.A.}},
\bauthor{\bsnm{Arif}, \binits{M.}}:
\batitle{The digital duo: Exploring the impact of ai chatbots and digital marketing strategies on consumer purchase intentions in pakistan’s e-commerce sector}.
\bjtitle{Journal for Social Science Archives}
\bvolume{3}(\bissue{1}),
\bfpage{265}--\blpage{286}
(\byear{2025})
\end{barticle}
\endbibitem

\bibitem[\protect\citeauthoryear{Aldayel and Magdy}{2022}]{aldayel2022characterizing}
\begin{barticle}
\bauthor{\bsnm{Aldayel}, \binits{A.}},
\bauthor{\bsnm{Magdy}, \binits{W.}}:
\batitle{Characterizing the role of bots’ in polarized stance on social media}.
\bjtitle{Social Network Analysis and Mining}
\bvolume{12}(\bissue{1}),
\bfpage{30}
(\byear{2022})
\end{barticle}
\endbibitem

\bibitem[\protect\citeauthoryear{Ng et~al.}{2024}]{ng2024cyborgs}
\begin{barticle}
\bauthor{\bsnm{Ng}, \binits{L.H.X.}},
\bauthor{\bsnm{Robertson}, \binits{D.C.}},
\bauthor{\bsnm{Carley}, \binits{K.M.}}:
\batitle{Cyborgs for strategic communication on social media}.
\bjtitle{Big Data \& Society}
\bvolume{11}(\bissue{1}),
\bfpage{20539517241231275}
(\byear{2024})
\end{barticle}
\endbibitem

\bibitem[\protect\citeauthoryear{Ng and Carley}{2024}]{ng2024assembling}
\begin{barticle}
\bauthor{\bsnm{Ng}, \binits{L.H.X.}},
\bauthor{\bsnm{Carley}, \binits{K.M.}}:
\batitle{Assembling a multi-platform ensemble social bot detector with applications to us 2020 elections}.
\bjtitle{Social Network Analysis and Mining}
\bvolume{14}(\bissue{1}),
\bfpage{45}
(\byear{2024})
\end{barticle}
\endbibitem

\bibitem[\protect\citeauthoryear{A{\"\i}meur et~al.}{2023}]{aimeur2023fake}
\begin{barticle}
\bauthor{\bsnm{A{\"\i}meur}, \binits{E.}},
\bauthor{\bsnm{Amri}, \binits{S.}},
\bauthor{\bsnm{Brassard}, \binits{G.}}:
\batitle{Fake news, disinformation and misinformation in social media: a review}.
\bjtitle{Social Network Analysis and Mining}
\bvolume{13}(\bissue{1}),
\bfpage{30}
(\byear{2023})
\end{barticle}
\endbibitem

\bibitem[\protect\citeauthoryear{Freelon et~al.}{2022}]{freelon2022black}
\begin{barticle}
\bauthor{\bsnm{Freelon}, \binits{D.}},
\bauthor{\bsnm{Bossetta}, \binits{M.}},
\bauthor{\bsnm{Wells}, \binits{C.}},
\bauthor{\bsnm{Lukito}, \binits{J.}},
\bauthor{\bsnm{Xia}, \binits{Y.}},
\bauthor{\bsnm{Adams}, \binits{K.}}:
\batitle{Black trolls matter: Racial and ideological asymmetries in social media disinformation}.
\bjtitle{Social Science Computer Review}
\bvolume{40}(\bissue{3}),
\bfpage{560}--\blpage{578}
(\byear{2022})
\end{barticle}
\endbibitem

\bibitem[\protect\citeauthoryear{Lu and Lee}{2024}]{lu2024agents}
\begin{botherref}
\oauthor{\bsnm{Lu}, \binits{H.-C.}},
\oauthor{\bsnm{Lee}, \binits{H.-w.}}:
Agents of discord: Modeling the impact of political bots on opinion polarization in social networks.
Social Science Computer Review,
08944393241270382
(2024)
\end{botherref}
\endbibitem

\bibitem[\protect\citeauthoryear{Marigliano et~al.}{2024}]{marigliano2024analyzing}
\begin{barticle}
\bauthor{\bsnm{Marigliano}, \binits{R.}},
\bauthor{\bsnm{Ng}, \binits{L.H.X.}},
\bauthor{\bsnm{Carley}, \binits{K.M.}}:
\batitle{Analyzing digital propaganda and conflict rhetoric: a study on russia’s bot-driven campaigns and counter-narratives during the ukraine crisis}.
\bjtitle{Social Network Analysis and Mining}
\bvolume{14}(\bissue{1}),
\bfpage{170}
(\byear{2024})
\end{barticle}
\endbibitem

\bibitem[\protect\citeauthoryear{Tardelli et~al.}{2024}]{tardelli2024multifaceted}
\begin{barticle}
\bauthor{\bsnm{Tardelli}, \binits{S.}},
\bauthor{\bsnm{Nizzoli}, \binits{L.}},
\bauthor{\bsnm{Avvenuti}, \binits{M.}},
\bauthor{\bsnm{Cresci}, \binits{S.}},
\bauthor{\bsnm{Tesconi}, \binits{M.}}:
\batitle{Multifaceted online coordinated behavior in the 2020 us presidential election}.
\bjtitle{EPJ Data Science}
\bvolume{13}(\bissue{1}),
\bfpage{33}
(\byear{2024})
\end{barticle}
\endbibitem

\bibitem[\protect\citeauthoryear{Blane et~al.}{2023}]{blane2023analyzing}
\begin{bchapter}
\bauthor{\bsnm{Blane}, \binits{J.T.}},
\bauthor{\bsnm{Ng}, \binits{L.H.X.}},
\bauthor{\bsnm{Carley}, \binits{K.M.}}:
\bctitle{Analyzing social-cyber maneuvers for spreading covid-19 pro-and anti-vaccine information}.
In: \bbtitle{Vaccine Communication Online: Counteracting Misinformation, Rumors and Lies},
pp. \bfpage{57}--\blpage{80}.
\bpublisher{Springer}, \blocation{???}
(\byear{2023})
\end{bchapter}
\endbibitem

\bibitem[\protect\citeauthoryear{Diab et~al.}{2023}]{diab2023online}
\begin{bchapter}
\bauthor{\bsnm{Diab}, \binits{A.}},
\bauthor{\bsnm{Jagdagdorj}, \binits{B.-E.}},
\bauthor{\bsnm{Ng}, \binits{L.H.X.}},
\bauthor{\bsnm{Lin}, \binits{Y.-R.}},
\bauthor{\bsnm{Yoder}, \binits{M.M.}}:
\bctitle{Online to offline crossover of white supremacist propaganda}.
In: \bbtitle{Companion Proceedings of the ACM Web Conference 2023},
pp. \bfpage{1308}--\blpage{1316}
(\byear{2023})
\end{bchapter}
\endbibitem

\bibitem[\protect\citeauthoryear{Woolley and Howard}{2016}]{woolley2016social}
\begin{bchapter}
\bauthor{\bsnm{Woolley}, \binits{S.C.}},
\bauthor{\bsnm{Howard}, \binits{P.N.}}:
\bctitle{Social media, revolution, and the rise of the political bot}.
In: \bbtitle{Routledge Handbook of Media, Conflict and Security},
pp. \bfpage{302}--\blpage{312}.
\bpublisher{Routledge}, \blocation{???}
(\byear{2016})
\end{bchapter}
\endbibitem

\bibitem[\protect\citeauthoryear{Carley}{2020}]{carley2020social}
\begin{barticle}
\bauthor{\bsnm{Carley}, \binits{K.M.}}:
\batitle{Social cybersecurity: an emerging science}.
\bjtitle{Computational and mathematical organization theory}
\bvolume{26}(\bissue{4}),
\bfpage{365}--\blpage{381}
(\byear{2020})
\end{barticle}
\endbibitem

\bibitem[\protect\citeauthoryear{Nakov et~al.}{2021}]{nakov2021automated}
\begin{botherref}
\oauthor{\bsnm{Nakov}, \binits{P.}},
\oauthor{\bsnm{Corney}, \binits{D.}},
\oauthor{\bsnm{Hasanain}, \binits{M.}},
\oauthor{\bsnm{Alam}, \binits{F.}},
\oauthor{\bsnm{Elsayed}, \binits{T.}},
\oauthor{\bsnm{Barr{\'o}n-Cede{\~n}o}, \binits{A.}},
\oauthor{\bsnm{Papotti}, \binits{P.}},
\oauthor{\bsnm{Shaar}, \binits{S.}},
\oauthor{\bsnm{Martino}, \binits{G.D.S.}}:
Automated fact-checking for assisting human fact-checkers.
arXiv preprint arXiv:2103.07769
(2021)
\end{botherref}
\endbibitem

\bibitem[\protect\citeauthoryear{He et~al.}{2024}]{he2024platform}
\begin{botherref}
\oauthor{\bsnm{He}, \binits{Q.}},
\oauthor{\bsnm{Hong}, \binits{Y.}},
\oauthor{\bsnm{Raghu}, \binits{T.}}:
Platform governance with algorithm-based content moderation: An empirical study on reddit.
Information Systems Research
(2024)
\end{botherref}
\endbibitem

\bibitem[\protect\citeauthoryear{O'Hara}{2022}]{o2022automated}
\begin{barticle}
\bauthor{\bsnm{O'Hara}, \binits{I.}}:
\batitle{Automated epistemology: Bots, computational propaganda \& information literacy instruction}.
\bjtitle{The Journal of Academic Librarianship}
\bvolume{48}(\bissue{4}),
\bfpage{102540}
(\byear{2022})
\end{barticle}
\endbibitem

\bibitem[\protect\citeauthoryear{Al-Rawi and Shukla}{2020}]{al2020bots}
\begin{barticle}
\bauthor{\bsnm{Al-Rawi}, \binits{A.}},
\bauthor{\bsnm{Shukla}, \binits{V.}}:
\batitle{Bots as active news promoters: A digital analysis of covid-19 tweets}.
\bjtitle{Information}
\bvolume{11}(\bissue{10}),
\bfpage{461}
(\byear{2020})
\end{barticle}
\endbibitem

\bibitem[\protect\citeauthoryear{Ng and Carley}{2021}]{ng2021bot}
\begin{bchapter}
\bauthor{\bsnm{Ng}, \binits{L.H.X.}},
\bauthor{\bsnm{Carley}, \binits{K.M.}}:
\bctitle{Bot-based emotion behavior differences in images during kashmir black day event}.
In: \bbtitle{International Conference on Social Computing, Behavioral-Cultural Modeling and Prediction and Behavior Representation in Modeling and Simulation},
pp. \bfpage{184}--\blpage{194}
(\byear{2021}).
\bcomment{Springer}
\end{bchapter}
\endbibitem

\bibitem[\protect\citeauthoryear{Beskow and Carley}{2018}]{beskow2018bot}
\begin{bchapter}
\bauthor{\bsnm{Beskow}, \binits{D.M.}},
\bauthor{\bsnm{Carley}, \binits{K.M.}}:
\bctitle{Bot-hunter: a tiered approach to detecting \& characterizing automated activity on twitter}.
In: \bbtitle{Conference Paper. SBP-BRiMS: International Conference on Social Computing, Behavioral-cultural Modeling and Prediction and Behavior Representation in Modeling and Simulation},
vol. \bseriesno{3}
(\byear{2018})
\end{bchapter}
\endbibitem

\bibitem[\protect\citeauthoryear{Shao et~al.}{2018}]{shao2018spread}
\begin{barticle}
\bauthor{\bsnm{Shao}, \binits{C.}},
\bauthor{\bsnm{Ciampaglia}, \binits{G.L.}},
\bauthor{\bsnm{Varol}, \binits{O.}},
\bauthor{\bsnm{Yang}, \binits{K.-C.}},
\bauthor{\bsnm{Flammini}, \binits{A.}},
\bauthor{\bsnm{Menczer}, \binits{F.}}:
\batitle{The spread of low-credibility content by social bots}.
\bjtitle{Nature communications}
\bvolume{9}(\bissue{1}),
\bfpage{1}--\blpage{9}
(\byear{2018})
\end{barticle}
\endbibitem

\bibitem[\protect\citeauthoryear{Ng and Carley}{2022}]{ng2022pro}
\begin{barticle}
\bauthor{\bsnm{Ng}, \binits{L.H.X.}},
\bauthor{\bsnm{Carley}, \binits{K.M.}}:
\batitle{Pro or anti? a social influence model of online stance flipping}.
\bjtitle{IEEE Transactions on Network Science and Engineering}
\bvolume{10}(\bissue{1}),
\bfpage{3}--\blpage{19}
(\byear{2022})
\end{barticle}
\endbibitem

\bibitem[\protect\citeauthoryear{Gambini et~al.}{2024}]{gambini2024anatomy}
\begin{barticle}
\bauthor{\bsnm{Gambini}, \binits{M.}},
\bauthor{\bsnm{Tardelli}, \binits{S.}},
\bauthor{\bsnm{Tesconi}, \binits{M.}}:
\batitle{The anatomy of conspiracy theorists: unveiling traits using a comprehensive twitter dataset}.
\bjtitle{Computer Communications}
\bvolume{217},
\bfpage{25}--\blpage{40}
(\byear{2024})
\end{barticle}
\endbibitem

\bibitem[\protect\citeauthoryear{Tsvetkova et~al.}{2017}]{tsvetkova2017even}
\begin{barticle}
\bauthor{\bsnm{Tsvetkova}, \binits{M.}},
\bauthor{\bsnm{Garc{\'\i}a-Gavilanes}, \binits{R.}},
\bauthor{\bsnm{Floridi}, \binits{L.}},
\bauthor{\bsnm{Yasseri}, \binits{T.}}:
\batitle{Even good bots fight: The case of wikipedia}.
\bjtitle{PloS one}
\bvolume{12}(\bissue{2}),
\bfpage{0171774}
(\byear{2017})
\end{barticle}
\endbibitem

\bibitem[\protect\citeauthoryear{Ferrara et~al.}{2016}]{ferrara2016rise}
\begin{barticle}
\bauthor{\bsnm{Ferrara}, \binits{E.}},
\bauthor{\bsnm{Varol}, \binits{O.}},
\bauthor{\bsnm{Davis}, \binits{C.}},
\bauthor{\bsnm{Menczer}, \binits{F.}},
\bauthor{\bsnm{Flammini}, \binits{A.}}:
\batitle{The rise of social bots}.
\bjtitle{Communications of the ACM}
\bvolume{59}(\bissue{7}),
\bfpage{96}--\blpage{104}
(\byear{2016})
\end{barticle}
\endbibitem

\bibitem[\protect\citeauthoryear{Davis et~al.}{2016}]{davis2016botornot}
\begin{bchapter}
\bauthor{\bsnm{Davis}, \binits{C.A.}},
\bauthor{\bsnm{Varol}, \binits{O.}},
\bauthor{\bsnm{Ferrara}, \binits{E.}},
\bauthor{\bsnm{Flammini}, \binits{A.}},
\bauthor{\bsnm{Menczer}, \binits{F.}}:
\bctitle{Botornot: A system to evaluate social bots}.
In: \bbtitle{Proceedings of the 25th International Conference Companion on World Wide Web},
pp. \bfpage{273}--\blpage{274}
(\byear{2016})
\end{bchapter}
\endbibitem

\bibitem[\protect\citeauthoryear{Chavoshi et~al.}{2016}]{chavoshi2016debot}
\begin{bchapter}
\bauthor{\bsnm{Chavoshi}, \binits{N.}},
\bauthor{\bsnm{Hamooni}, \binits{H.}},
\bauthor{\bsnm{Mueen}, \binits{A.}}:
\bctitle{Debot: Twitter bot detection via warped correlation.}
In: \bbtitle{Icdm},
vol. \bseriesno{18},
pp. \bfpage{28}--\blpage{65}
(\byear{2016})
\end{bchapter}
\endbibitem

\bibitem[\protect\citeauthoryear{Sayyadiharikandeh et~al.}{2020}]{10.1145/3340531.3412698}
\begin{bchapter}
\bauthor{\bsnm{Sayyadiharikandeh}, \binits{M.}},
\bauthor{\bsnm{Varol}, \binits{O.}},
\bauthor{\bsnm{Yang}, \binits{K.-C.}},
\bauthor{\bsnm{Flammini}, \binits{A.}},
\bauthor{\bsnm{Menczer}, \binits{F.}}:
\bctitle{Detection of novel social bots by ensembles of specialized classifiers}.
In: \bbtitle{Proceedings of the 29th ACM International Conference on Information \& Knowledge Management}.
\bsertitle{CIKM '20},
pp. \bfpage{2725}--\blpage{2732}.
\bpublisher{Association for Computing Machinery},
\blocation{New York, NY, USA}
(\byear{2020}).
\doiurl{10.1145/3340531.3412698} .
\burl{https://doi.org/10.1145/3340531.3412698}
\end{bchapter}
\endbibitem

\bibitem[\protect\citeauthoryear{Ng and Carley}{2023}]{ng2023botbuster}
\begin{bchapter}
\bauthor{\bsnm{Ng}, \binits{L.H.X.}},
\bauthor{\bsnm{Carley}, \binits{K.M.}}:
\bctitle{Botbuster: Multi-platform bot detection using a mixture of experts}.
In: \bbtitle{Proceedings of the International AAAI Conference on Web and Social Media},
vol. \bseriesno{17},
pp. \bfpage{686}--\blpage{697}
(\byear{2023})
\end{bchapter}
\endbibitem

\bibitem[\protect\citeauthoryear{Feng et~al.}{2022}]{feng2022twibot}
\begin{barticle}
\bauthor{\bsnm{Feng}, \binits{S.}},
\bauthor{\bsnm{Tan}, \binits{Z.}},
\bauthor{\bsnm{Wan}, \binits{H.}},
\bauthor{\bsnm{Wang}, \binits{N.}},
\bauthor{\bsnm{Chen}, \binits{Z.}},
\bauthor{\bsnm{Zhang}, \binits{B.}},
\bauthor{\bsnm{Zheng}, \binits{Q.}},
\bauthor{\bsnm{Zhang}, \binits{W.}},
\bauthor{\bsnm{Lei}, \binits{Z.}},
\bauthor{\bsnm{Yang}, \binits{S.}}, \betal:
\batitle{Twibot-22: Towards graph-based twitter bot detection}.
\bjtitle{Advances in Neural Information Processing Systems}
\bvolume{35},
\bfpage{35254}--\blpage{35269}
(\byear{2022})
\end{barticle}
\endbibitem

\bibitem[\protect\citeauthoryear{McBride et~al.}{2020}]{mcbride2020social}
\begin{botherref}
\oauthor{\bsnm{McBride}, \binits{M.K.}},
\oauthor{\bsnm{Gold}, \binits{Z.}},
\oauthor{\bsnm{Stricklin}, \binits{K.}}:
Social media bots: Implications for special operations forces.
Center for Naval Analysis, September
(2020)
\end{botherref}
\endbibitem

\bibitem[\protect\citeauthoryear{Oentaryo et~al.}{2016}]{oentaryo2016profiling}
\begin{bchapter}
\bauthor{\bsnm{Oentaryo}, \binits{R.J.}},
\bauthor{\bsnm{Murdopo}, \binits{A.}},
\bauthor{\bsnm{Prasetyo}, \binits{P.K.}},
\bauthor{\bsnm{Lim}, \binits{E.-P.}}:
\bctitle{On profiling bots in social media}.
In: \bbtitle{Social Informatics: 8th International Conference, SocInfo 2016, Bellevue, WA, USA, November 11-14, 2016, Proceedings, Part I 8},
pp. \bfpage{92}--\blpage{109}
(\byear{2016}).
\bcomment{Springer}
\end{bchapter}
\endbibitem

\bibitem[\protect\citeauthoryear{Abokhodair et~al.}{2015}]{abokhodair2015dissecting}
\begin{bchapter}
\bauthor{\bsnm{Abokhodair}, \binits{N.}},
\bauthor{\bsnm{Yoo}, \binits{D.}},
\bauthor{\bsnm{McDonald}, \binits{D.W.}}:
\bctitle{Dissecting a social botnet: Growth, content and influence in twitter}.
In: \bbtitle{Proceedings of the 18th ACM Conference on Computer Supported Cooperative Work \& Social Computing},
pp. \bfpage{839}--\blpage{851}
(\byear{2015})
\end{bchapter}
\endbibitem

\bibitem[\protect\citeauthoryear{Elmas et~al.}{2022}]{elmas2022characterizing}
\begin{bchapter}
\bauthor{\bsnm{Elmas}, \binits{T.}},
\bauthor{\bsnm{Overdorf}, \binits{R.}},
\bauthor{\bsnm{Aberer}, \binits{K.}}:
\bctitle{Characterizing retweet bots: The case of black market accounts}.
In: \bbtitle{Proceedings of the International AAAI Conference on Web and Social Media},
vol. \bseriesno{16},
pp. \bfpage{171}--\blpage{182}
(\byear{2022})
\end{bchapter}
\endbibitem

\bibitem[\protect\citeauthoryear{Chu et~al.}{2012}]{chu2012detecting}
\begin{bchapter}
\bauthor{\bsnm{Chu}, \binits{Z.}},
\bauthor{\bsnm{Widjaja}, \binits{I.}},
\bauthor{\bsnm{Wang}, \binits{H.}}:
\bctitle{Detecting social spam campaigns on twitter}.
In: \bbtitle{Applied Cryptography and Network Security: 10th International Conference, ACNS 2012, Singapore, June 26-29, 2012. Proceedings 10},
pp. \bfpage{455}--\blpage{472}
(\byear{2012}).
\bcomment{Springer}
\end{bchapter}
\endbibitem

\bibitem[\protect\citeauthoryear{Jamison et~al.}{2019}]{jamison2019malicious}
\begin{barticle}
\bauthor{\bsnm{Jamison}, \binits{A.M.}},
\bauthor{\bsnm{Broniatowski}, \binits{D.A.}},
\bauthor{\bsnm{Quinn}, \binits{S.C.}}:
\batitle{Malicious actors on twitter: A guide for public health researchers}.
\bjtitle{American journal of public health}
\bvolume{109}(\bissue{5}),
\bfpage{688}--\blpage{692}
(\byear{2019})
\end{barticle}
\endbibitem

\bibitem[\protect\citeauthoryear{Lee et~al.}{2011}]{lee2011seven}
\begin{bchapter}
\bauthor{\bsnm{Lee}, \binits{K.}},
\bauthor{\bsnm{Eoff}, \binits{B.}},
\bauthor{\bsnm{Caverlee}, \binits{J.}}:
\bctitle{Seven months with the devils: A long-term study of content polluters on twitter}.
In: \bbtitle{Proceedings of the International AAAI Conference on Web and Social Media},
vol. \bseriesno{5},
pp. \bfpage{185}--\blpage{192}
(\byear{2011})
\end{bchapter}
\endbibitem

\bibitem[\protect\citeauthoryear{Mbona and Eloff}{2023}]{mbona2023classifying}
\begin{barticle}
\bauthor{\bsnm{Mbona}, \binits{I.}},
\bauthor{\bsnm{Eloff}, \binits{J.H.}}:
\batitle{Classifying social media bots as malicious or benign using semi-supervised machine learning}.
\bjtitle{Journal of Cybersecurity}
\bvolume{9}(\bissue{1}),
\bfpage{015}
(\byear{2023})
\end{barticle}
\endbibitem

\bibitem[\protect\citeauthoryear{Ng et~al.}{2025}]{ng2025aurasight}
\begin{botherref}
\oauthor{\bsnm{Ng}, \binits{L.H.X.}},
\oauthor{\bsnm{Kang}, \binits{B.N.Y.}},
\oauthor{\bsnm{Carley}, \binits{K.M.}}:
Aurasight: Generating realistic social media data.
Technical Report CMU-S3D-25-109,
Carnegie Mellon University
(2025).
\url{http://reports-archive.adm.cs.cmu.edu/anon/anon/home/ftp/s3d2025/CMU-S3D-25-109.pdf}
\end{botherref}
\endbibitem

\bibitem[\protect\citeauthoryear{Ferrara et~al.}{2020}]{ferrara2020characterizing}
\begin{botherref}
\oauthor{\bsnm{Ferrara}, \binits{E.}},
\oauthor{\bsnm{Chang}, \binits{H.}},
\oauthor{\bsnm{Chen}, \binits{E.}},
\oauthor{\bsnm{Muric}, \binits{G.}},
\oauthor{\bsnm{Patel}, \binits{J.}}:
Characterizing social media manipulation in the 2020 us presidential election.
First Monday
(2020)
\end{botherref}
\endbibitem

\bibitem[\protect\citeauthoryear{Avvenuti et~al.}{2014}]{avvenuti2014ears}
\begin{bchapter}
\bauthor{\bsnm{Avvenuti}, \binits{M.}},
\bauthor{\bsnm{Cresci}, \binits{S.}},
\bauthor{\bsnm{Marchetti}, \binits{A.}},
\bauthor{\bsnm{Meletti}, \binits{C.}},
\bauthor{\bsnm{Tesconi}, \binits{M.}}:
\bctitle{Ears (earthquake alert and report system) a real time decision support system for earthquake crisis management}.
In: \bbtitle{Proceedings of the 20th ACM SIGKDD International Conference on Knowledge Discovery and Data Mining},
pp. \bfpage{1749}--\blpage{1758}
(\byear{2014})
\end{bchapter}
\endbibitem

\bibitem[\protect\citeauthoryear{Crooks et~al.}{2013}]{crooks2013earthquake}
\begin{barticle}
\bauthor{\bsnm{Crooks}, \binits{A.}},
\bauthor{\bsnm{Croitoru}, \binits{A.}},
\bauthor{\bsnm{Stefanidis}, \binits{A.}},
\bauthor{\bsnm{Radzikowski}, \binits{J.}}:
\batitle{\# earthquake: Twitter as a distributed sensor system}.
\bjtitle{Transactions in GIS}
\bvolume{17}(\bissue{1}),
\bfpage{124}--\blpage{147}
(\byear{2013})
\end{barticle}
\endbibitem

\bibitem[\protect\citeauthoryear{Hofeditz et~al.}{2019}]{hofeditz2019meaningful}
\begin{bchapter}
\bauthor{\bsnm{Hofeditz}, \binits{L.}},
\bauthor{\bsnm{Ehnis}, \binits{C.}},
\bauthor{\bsnm{Bunker}, \binits{D.}},
\bauthor{\bsnm{Brachten}, \binits{F.}},
\bauthor{\bsnm{Stieglitz}, \binits{S.}}:
\bctitle{Meaningful use of social bots? possible applications in crisis communication during disasters.}
In: \bbtitle{ECIS},
pp. \bfpage{1}--\blpage{16}
(\byear{2019})
\end{bchapter}
\endbibitem

\bibitem[\protect\citeauthoryear{Ng et~al.}{2024}]{ng2024tiny}
\begin{bchapter}
\bauthor{\bsnm{Ng}, \binits{L.H.X.}},
\bauthor{\bsnm{Bartulovic}, \binits{M.}},
\bauthor{\bsnm{Carley}, \binits{K.M.}}:
\bctitle{Tiny-botbuster: Identifying automated political coordination in digital campaigns}.
In: \bbtitle{International Conference on Social Computing, Behavioral-Cultural Modeling and Prediction and Behavior Representation in Modeling and Simulation},
pp. \bfpage{25}--\blpage{34}
(\byear{2024}).
\bcomment{Springer}
\end{bchapter}
\endbibitem

\bibitem[\protect\citeauthoryear{Woolley}{2016}]{woolley2016automating}
\begin{botherref}
\oauthor{\bsnm{Woolley}, \binits{S.C.}}:
Automating power: Social bot interference in global politics.
First Monday
(2016)
\end{botherref}
\endbibitem

\bibitem[\protect\citeauthoryear{Jacobs et~al.}{2023}]{jacobs2023tracking}
\begin{bchapter}
\bauthor{\bsnm{Jacobs}, \binits{C.S.}},
\bauthor{\bsnm{Ng}, \binits{L.H.X.}},
\bauthor{\bsnm{Carley}, \binits{K.M.}}:
\bctitle{Tracking china’s cross-strait bot networks against taiwan}.
In: \bbtitle{International Conference on Social Computing, Behavioral-cultural Modeling and Prediction and Behavior Representation in Modeling and Simulation},
pp. \bfpage{115}--\blpage{125}
(\byear{2023}).
\bcomment{Springer}
\end{bchapter}
\endbibitem

\bibitem[\protect\citeauthoryear{Ng and Carley}{2023}]{ng2023combined}
\begin{barticle}
\bauthor{\bsnm{Ng}, \binits{L.H.X.}},
\bauthor{\bsnm{Carley}, \binits{K.M.}}:
\batitle{A combined synchronization index for evaluating collective action social media}.
\bjtitle{Applied network science}
\bvolume{8}(\bissue{1}),
\bfpage{1}
(\byear{2023})
\end{barticle}
\endbibitem

\bibitem[\protect\citeauthoryear{Badawy et~al.}{2018}]{badawy2018analyzing}
\begin{bchapter}
\bauthor{\bsnm{Badawy}, \binits{A.}},
\bauthor{\bsnm{Ferrara}, \binits{E.}},
\bauthor{\bsnm{Lerman}, \binits{K.}}:
\bctitle{Analyzing the digital traces of political manipulation: The 2016 russian interference twitter campaign}.
In: \bbtitle{2018 IEEE/ACM International Conference on Advances in Social Networks Analysis and Mining (ASONAM)},
pp. \bfpage{258}--\blpage{265}
(\byear{2018}).
\bcomment{IEEE}
\end{bchapter}
\endbibitem

\bibitem[\protect\citeauthoryear{Elmas}{2022}]{elmas2022role}
\begin{botherref}
\oauthor{\bsnm{Elmas}, \binits{T.}}:
The role of compromised accounts in social media manipulation.
PhD thesis,
EPFL
(2022)
\end{botherref}
\endbibitem

\bibitem[\protect\citeauthoryear{Chen et~al.}{2022}]{chen2022social}
\begin{barticle}
\bauthor{\bsnm{Chen}, \binits{L.}},
\bauthor{\bsnm{Chen}, \binits{J.}},
\bauthor{\bsnm{Xia}, \binits{C.}}:
\batitle{Social network behavior and public opinion manipulation}.
\bjtitle{Journal of Information Security and Applications}
\bvolume{64},
\bfpage{103060}
(\byear{2022})
\end{barticle}
\endbibitem

\bibitem[\protect\citeauthoryear{Danaditya et~al.}{2022}]{danaditya2022curious}
\begin{barticle}
\bauthor{\bsnm{Danaditya}, \binits{A.}},
\bauthor{\bsnm{Ng}, \binits{L.H.X.}},
\bauthor{\bsnm{Carley}, \binits{K.M.}}:
\batitle{From curious hashtags to polarized effect: profiling coordinated actions in indonesian twitter discourse}.
\bjtitle{Social Network Analysis and Mining}
\bvolume{12}(\bissue{1}),
\bfpage{105}
(\byear{2022})
\end{barticle}
\endbibitem

\bibitem[\protect\citeauthoryear{McKelvey and Dubois}{2017}]{mckelvey2017computational}
\begin{botherref}
\oauthor{\bsnm{McKelvey}, \binits{F.}},
\oauthor{\bsnm{Dubois}, \binits{E.}}:
Computational propaganda in canada: The use of political bots
(2017)
\end{botherref}
\endbibitem

\bibitem[\protect\citeauthoryear{Duan et~al.}{2022}]{duan2022algorithmic}
\begin{barticle}
\bauthor{\bsnm{Duan}, \binits{Z.}},
\bauthor{\bsnm{Li}, \binits{J.}},
\bauthor{\bsnm{Lukito}, \binits{J.}},
\bauthor{\bsnm{Yang}, \binits{K.-C.}},
\bauthor{\bsnm{Chen}, \binits{F.}},
\bauthor{\bsnm{Shah}, \binits{D.V.}},
\bauthor{\bsnm{Yang}, \binits{S.}}:
\batitle{Algorithmic agents in the hybrid media system: Social bots, selective amplification, and partisan news about covid-19}.
\bjtitle{Human Communication Research}
\bvolume{48}(\bissue{3}),
\bfpage{516}--\blpage{542}
(\byear{2022})
\end{barticle}
\endbibitem

\bibitem[\protect\citeauthoryear{Magelinski et~al.}{2022}]{magelinski2022synchronized}
\begin{botherref}
\oauthor{\bsnm{Magelinski}, \binits{T.}},
\oauthor{\bsnm{Ng}, \binits{L.}},
\oauthor{\bsnm{Carley}, \binits{K.}}:
A synchronized action framework for detection of coordination on social media.
Journal of Online Trust and Safety
\textbf{1}(2)
(2022)
\end{botherref}
\endbibitem

\bibitem[\protect\citeauthoryear{Uyheng and Carley}{2020}]{uyheng2020bot}
\begin{bchapter}
\bauthor{\bsnm{Uyheng}, \binits{J.}},
\bauthor{\bsnm{Carley}, \binits{K.M.}}:
\bctitle{Bot impacts on public sentiment and community structures: Comparative analysis of three elections in the asia-pacific}.
In: \bbtitle{Social, Cultural, and Behavioral Modeling: 13th International Conference, SBP-BRiMS 2020, Washington, DC, USA, October 18--21, 2020, Proceedings 13},
pp. \bfpage{12}--\blpage{22}
(\byear{2020}).
\bcomment{Springer}
\end{bchapter}
\endbibitem

\bibitem[\protect\citeauthoryear{Sindhu and Bharti}{2024}]{sindhu2024influence}
\begin{barticle}
\bauthor{\bsnm{Sindhu}, \binits{P.}},
\bauthor{\bsnm{Bharti}, \binits{K.}}:
\batitle{Influence of chatbots on purchase intention in social commerce}.
\bjtitle{Behaviour \& Information Technology}
\bvolume{43}(\bissue{2}),
\bfpage{331}--\blpage{352}
(\byear{2024})
\end{barticle}
\endbibitem

\bibitem[\protect\citeauthoryear{Chang and Ferrara}{2022}]{chang2022comparative}
\begin{barticle}
\bauthor{\bsnm{Chang}, \binits{H.-C.H.}},
\bauthor{\bsnm{Ferrara}, \binits{E.}}:
\batitle{Comparative analysis of social bots and humans during the covid-19 pandemic}.
\bjtitle{Journal of Computational Social Science}
\bvolume{5}(\bissue{2}),
\bfpage{1409}--\blpage{1425}
(\byear{2022})
\end{barticle}
\endbibitem

\bibitem[\protect\citeauthoryear{Gorwa and Guilbeault}{2020}]{gorwa2020unpacking}
\begin{barticle}
\bauthor{\bsnm{Gorwa}, \binits{R.}},
\bauthor{\bsnm{Guilbeault}, \binits{D.}}:
\batitle{Unpacking the social media bot: A typology to guide research and policy}.
\bjtitle{Policy \& Internet}
\bvolume{12}(\bissue{2}),
\bfpage{225}--\blpage{248}
(\byear{2020})
\end{barticle}
\endbibitem

\bibitem[\protect\citeauthoryear{Lotan et~al.}{2011}]{lotan2011arab}
\begin{barticle}
\bauthor{\bsnm{Lotan}, \binits{G.}},
\bauthor{\bsnm{Graeff}, \binits{E.}},
\bauthor{\bsnm{Ananny}, \binits{M.}},
\bauthor{\bsnm{Gaffney}, \binits{D.}},
\bauthor{\bsnm{Pearce}, \binits{I.}}, \betal:
\batitle{The arab spring| the revolutions were tweeted: Information flows during the 2011 tunisian and egyptian revolutions}.
\bjtitle{International journal of communication}
\bvolume{5},
\bfpage{31}
(\byear{2011})
\end{barticle}
\endbibitem

\bibitem[\protect\citeauthoryear{Liu et~al.}{2017}]{liu2017investigation}
\begin{barticle}
\bauthor{\bsnm{Liu}, \binits{X.}},
\bauthor{\bsnm{Burns}, \binits{A.C.}},
\bauthor{\bsnm{Hou}, \binits{Y.}}:
\batitle{An investigation of brand-related user-generated content on twitter}.
\bjtitle{Journal of Advertising}
\bvolume{46}(\bissue{2}),
\bfpage{236}--\blpage{247}
(\byear{2017})
\end{barticle}
\endbibitem

\bibitem[\protect\citeauthoryear{Ng and Carley}{2023}]{ng2023you}
\begin{barticle}
\bauthor{\bsnm{Ng}, \binits{L.H.X.}},
\bauthor{\bsnm{Carley}, \binits{K.M.}}:
\batitle{Do you hear the people sing? comparison of synchronized url and narrative themes in 2020 and 2023 french protests}.
\bjtitle{Frontiers in big Data}
\bvolume{6},
\bfpage{1221744}
(\byear{2023})
\end{barticle}
\endbibitem

\bibitem[\protect\citeauthoryear{Khaund et~al.}{2021}]{khaund2021social}
\begin{barticle}
\bauthor{\bsnm{Khaund}, \binits{T.}},
\bauthor{\bsnm{Kirdemir}, \binits{B.}},
\bauthor{\bsnm{Agarwal}, \binits{N.}},
\bauthor{\bsnm{Liu}, \binits{H.}},
\bauthor{\bsnm{Morstatter}, \binits{F.}}:
\batitle{Social bots and their coordination during online campaigns: a survey}.
\bjtitle{IEEE Transactions on Computational Social Systems}
\bvolume{9}(\bissue{2}),
\bfpage{530}--\blpage{545}
(\byear{2021})
\end{barticle}
\endbibitem

\bibitem[\protect\citeauthoryear{Himelein-Wachowiak et~al.}{2021}]{himelein2021bots}
\begin{barticle}
\bauthor{\bsnm{Himelein-Wachowiak}, \binits{M.}},
\bauthor{\bsnm{Giorgi}, \binits{S.}},
\bauthor{\bsnm{Devoto}, \binits{A.}},
\bauthor{\bsnm{Rahman}, \binits{M.}},
\bauthor{\bsnm{Ungar}, \binits{L.}},
\bauthor{\bsnm{Schwartz}, \binits{H.A.}},
\bauthor{\bsnm{Epstein}, \binits{D.H.}},
\bauthor{\bsnm{Leggio}, \binits{L.}},
\bauthor{\bsnm{Curtis}, \binits{B.}}:
\batitle{Bots and misinformation spread on social media: Implications for covid-19}.
\bjtitle{Journal of medical Internet research}
\bvolume{23}(\bissue{5}),
\bfpage{26933}
(\byear{2021})
\end{barticle}
\endbibitem

\bibitem[\protect\citeauthoryear{Elyashar et~al.}{2014}]{elyashar2014guided}
\begin{barticle}
\bauthor{\bsnm{Elyashar}, \binits{A.}},
\bauthor{\bsnm{Fire}, \binits{M.}},
\bauthor{\bsnm{Kagan}, \binits{D.}},
\bauthor{\bsnm{Elovici}, \binits{Y.}}:
\batitle{Guided socialbots: Infiltrating the social networks of specific organizations’ employees}.
\bjtitle{Ai Communications}
\bvolume{29}(\bissue{1}),
\bfpage{87}--\blpage{106}
(\byear{2014})
\end{barticle}
\endbibitem

\bibitem[\protect\citeauthoryear{Arnaudo}{2017}]{arnaudo2017computational}
\begin{botherref}
\oauthor{\bsnm{Arnaudo}, \binits{D.}}:
Computational propaganda in brazil: Social bots during elections
(2017)
\end{botherref}
\endbibitem

\bibitem[\protect\citeauthoryear{Ferrara}{2020}]{ferrara2020covid}
\begin{botherref}
\oauthor{\bsnm{Ferrara}, \binits{E.}}:
\# covid-19 on twitter: Bots, conspiracies, and social media activism.
arXiv preprint arXiv: 2004.09531
(2020)
\end{botherref}
\endbibitem

\bibitem[\protect\citeauthoryear{Haustein et~al.}{2016}]{haustein2016tweets}
\begin{barticle}
\bauthor{\bsnm{Haustein}, \binits{S.}},
\bauthor{\bsnm{Bowman}, \binits{T.D.}},
\bauthor{\bsnm{Holmberg}, \binits{K.}},
\bauthor{\bsnm{Tsou}, \binits{A.}},
\bauthor{\bsnm{Sugimoto}, \binits{C.R.}},
\bauthor{\bsnm{Larivi{\`e}re}, \binits{V.}}:
\batitle{Tweets as impact indicators: Examining the implications of automated “bot” accounts on t witter}.
\bjtitle{Journal of the Association for Information Science and Technology}
\bvolume{67}(\bissue{1}),
\bfpage{232}--\blpage{238}
(\byear{2016})
\end{barticle}
\endbibitem

\bibitem[\protect\citeauthoryear{Jhaver et~al.}{2019}]{jhaver2019human}
\begin{barticle}
\bauthor{\bsnm{Jhaver}, \binits{S.}},
\bauthor{\bsnm{Birman}, \binits{I.}},
\bauthor{\bsnm{Gilbert}, \binits{E.}},
\bauthor{\bsnm{Bruckman}, \binits{A.}}:
\batitle{Human-machine collaboration for content regulation: The case of reddit automoderator}.
\bjtitle{ACM Transactions on Computer-Human Interaction (TOCHI)}
\bvolume{26}(\bissue{5}),
\bfpage{1}--\blpage{35}
(\byear{2019})
\end{barticle}
\endbibitem

\bibitem[\protect\citeauthoryear{{\"O}hman et~al.}{2019}]{ohman2019prayer}
\begin{barticle}
\bauthor{\bsnm{{\"O}hman}, \binits{C.}},
\bauthor{\bsnm{Gorwa}, \binits{R.}},
\bauthor{\bsnm{Floridi}, \binits{L.}}:
\batitle{Prayer-bots and religious worship on twitter: A call for a wider research agenda}.
\bjtitle{Minds and machines}
\bvolume{29}(\bissue{2}),
\bfpage{331}--\blpage{338}
(\byear{2019})
\end{barticle}
\endbibitem

\bibitem[\protect\citeauthoryear{Mantello et~al.}{2023}]{mantello2023automating}
\begin{bchapter}
\bauthor{\bsnm{Mantello}, \binits{P.}},
\bauthor{\bsnm{Ho}, \binits{T.M.}},
\bauthor{\bsnm{Podoletz}, \binits{L.}}:
\bctitle{Automating extremism: Mapping the affective roles of artificial agents in online radicalization}.
In: \bbtitle{The Palgrave Handbook of Malicious Use of AI and Psychological Security},
pp. \bfpage{81}--\blpage{103}.
\bpublisher{Springer}, \blocation{???}
(\byear{2023})
\end{bchapter}
\endbibitem

\bibitem[\protect\citeauthoryear{Benigni et~al.}{2017}]{benigni2017online}
\begin{barticle}
\bauthor{\bsnm{Benigni}, \binits{M.C.}},
\bauthor{\bsnm{Joseph}, \binits{K.}},
\bauthor{\bsnm{Carley}, \binits{K.M.}}:
\batitle{Online extremism and the communities that sustain it: Detecting the isis supporting community on twitter}.
\bjtitle{PloS one}
\bvolume{12}(\bissue{12}),
\bfpage{0181405}
(\byear{2017})
\end{barticle}
\endbibitem

\bibitem[\protect\citeauthoryear{Piccolo et~al.}{2021}]{piccolo2021chatbots}
\begin{bchapter}
\bauthor{\bsnm{Piccolo}, \binits{L.S.G.}},
\bauthor{\bsnm{Troullinou}, \binits{P.}},
\bauthor{\bsnm{Alani}, \binits{H.}}:
\bctitle{Chatbots to support children in coping with online threats: Socio-technical requirements}.
In: \bbtitle{Proceedings of the 2021 ACM Designing Interactive Systems Conference},
pp. \bfpage{1504}--\blpage{1517}
(\byear{2021})
\end{bchapter}
\endbibitem

\bibitem[\protect\citeauthoryear{Krueger and Osler}{2022}]{krueger2022communing}
\begin{barticle}
\bauthor{\bsnm{Krueger}, \binits{J.}},
\bauthor{\bsnm{Osler}, \binits{L.}}:
\batitle{Communing with the dead online: chatbots, grief, and continuing bonds}.
\bjtitle{Journal of Consciousness Studies}
\bvolume{29}(\bissue{9-10}),
\bfpage{222}--\blpage{252}
(\byear{2022})
\end{barticle}
\endbibitem

\bibitem[\protect\citeauthoryear{Moriuchi et~al.}{2021}]{moriuchi2021engagement}
\begin{barticle}
\bauthor{\bsnm{Moriuchi}, \binits{E.}},
\bauthor{\bsnm{Landers}, \binits{V.M.}},
\bauthor{\bsnm{Colton}, \binits{D.}},
\bauthor{\bsnm{Hair}, \binits{N.}}:
\batitle{Engagement with chatbots versus augmented reality interactive technology in e-commerce}.
\bjtitle{Journal of Strategic Marketing}
\bvolume{29}(\bissue{5}),
\bfpage{375}--\blpage{389}
(\byear{2021})
\end{barticle}
\endbibitem

\bibitem[\protect\citeauthoryear{Neff}{2016}]{neff2016talking}
\begin{botherref}
\oauthor{\bsnm{Neff}, \binits{G.}}:
Talking to bots: Symbiotic agency and the case of tay.
International Journal of Communication
(2016)
\end{botherref}
\endbibitem

\bibitem[\protect\citeauthoryear{Wolf et~al.}{2017}]{wolf2017we}
\begin{barticle}
\bauthor{\bsnm{Wolf}, \binits{M.J.}},
\bauthor{\bsnm{Miller}, \binits{K.}},
\bauthor{\bsnm{Grodzinsky}, \binits{F.S.}}:
\batitle{Why we should have seen that coming: comments on microsoft's tay" experiment," and wider implications}.
\bjtitle{Acm Sigcas Computers and Society}
\bvolume{47}(\bissue{3}),
\bfpage{54}--\blpage{64}
(\byear{2017})
\end{barticle}
\endbibitem

\bibitem[\protect\citeauthoryear{Veale et~al.}{2015}]{veale2015twitter}
\begin{bchapter}
\bauthor{\bsnm{Veale}, \binits{T.}},
\bauthor{\bsnm{Valitutti}, \binits{A.}},
\bauthor{\bsnm{Li}, \binits{G.}}:
\bctitle{Twitter: The best of bot worlds for automated wit}.
In: \bbtitle{Distributed, Ambient, and Pervasive Interactions: Third International Conference, DAPI 2015, Held as Part of HCI International 2015, Los Angeles, CA, USA, August 2-7, 2015, Proceedings 3},
pp. \bfpage{689}--\blpage{699}
(\byear{2015}).
\bcomment{Springer}
\end{bchapter}
\endbibitem

\bibitem[\protect\citeauthoryear{Elsholz et~al.}{2019}]{elsholz2019exploring}
\begin{bchapter}
\bauthor{\bsnm{Elsholz}, \binits{E.}},
\bauthor{\bsnm{Chamberlain}, \binits{J.}},
\bauthor{\bsnm{Kruschwitz}, \binits{U.}}:
\bctitle{Exploring language style in chatbots to increase perceived product value and user engagement}.
In: \bbtitle{Proceedings of the 2019 Conference on Human Information Interaction and Retrieval},
pp. \bfpage{301}--\blpage{305}
(\byear{2019})
\end{bchapter}
\endbibitem

\bibitem[\protect\citeauthoryear{Glenski et~al.}{2020}]{glenski2020user}
\begin{botherref}
\oauthor{\bsnm{Glenski}, \binits{M.}},
\oauthor{\bsnm{Volkova}, \binits{S.}},
\oauthor{\bsnm{Kumar}, \binits{S.}}:
User engagement with digital deception.
Disinformation, Misinformation, and Fake News in Social Media: Emerging Research Challenges and Opportunities,
39--61
(2020)
\end{botherref}
\endbibitem

\bibitem[\protect\citeauthoryear{Yeo}{2022}]{yeo2022models}
\begin{bchapter}
\bauthor{\bsnm{Yeo}, \binits{T.E.D.}}:
\bctitle{Models of viral propagation in digital contexts: How messages and ideas—from internet memes to fake news—created by consumers, bots, and marketers spread}.
In: \bbtitle{The Routledge Handbook of Digital Consumption},
pp. \bfpage{489}--\blpage{501}.
\bpublisher{Routledge}, \blocation{???}
(\byear{2022})
\end{bchapter}
\endbibitem

\bibitem[\protect\citeauthoryear{Hong and Oh}{2020}]{hong2020utilizing}
\begin{barticle}
\bauthor{\bsnm{Hong}, \binits{H.}},
\bauthor{\bsnm{Oh}, \binits{H.J.}}:
\batitle{Utilizing bots for sustainable news business: Understanding users’ perspectives of news bots in the age of social media}.
\bjtitle{Sustainability}
\bvolume{12}(\bissue{16}),
\bfpage{6515}
(\byear{2020})
\end{barticle}
\endbibitem

\bibitem[\protect\citeauthoryear{Lokot and Diakopoulos}{2016}]{lokot2016news}
\begin{barticle}
\bauthor{\bsnm{Lokot}, \binits{T.}},
\bauthor{\bsnm{Diakopoulos}, \binits{N.}}:
\batitle{News bots: Automating news and information dissemination on twitter}.
\bjtitle{Digital journalism}
\bvolume{4}(\bissue{6}),
\bfpage{682}--\blpage{699}
(\byear{2016})
\end{barticle}
\endbibitem

\end{thebibliography}

\end{document}